\documentstyle[12pt, twoside, epsf, here, psfig]{article}
\newcounter{tony}
\newcommand{\bb}{\mbox{\boldmath{$b$}}}
\newcommand{\bc}{\mbox{\boldmath{$c$}}}

\newcommand{\be}{\mbox{\boldmath{$e$}}}

\newcommand{\bg}{\mbox{\boldmath{$g$}}}

\newcommand{\bn}{\mbox{\boldmath{$n$}}}

\newcommand{\bv}{\mbox{\boldmath{$v$}}}

\newcommand{\bx}{\mbox{\boldmath{$x$}}}

\newcommand{\bD}{\mbox{\boldmath{$D$}}}

\newcommand{\bF}{\mbox{\boldmath{$F$}}}

\newcommand{\bH}{\mbox{\boldmath{$H$}}}
\newcommand{\bI}{\mbox{\boldmath{$I$}}}

\newcommand{\bP}{\mbox{\boldmath{$P$}}}

\newcommand{\btheta}{\mbox{\boldmath{$\theta$}}}

\newcommand{\bsigma}{\mbox{\boldmath{$\sigma$}}}

\newcommand{\btau}{\mbox{\boldmath{$\tau$}}}

\newcommand{\bomega}{\mbox{\boldmath{$\omega$}}}
\newcommand{\beq}{\begin{equation}}
\newcommand{\eeq}[1]{\label{eq:#1}\end{equation}}

\newcommand{\eqref}[1]{(\ref{eq:#1})}

      \newcommand{\beqn}{\begin{equation}}
      \newcommand{\eeqn}{\end{equation}}
      \newcommand{\beqna}{\begin{eqnarray}}
      \newcommand{\eeqna}{\end{eqnarray}}

\title{Incipient Motion Criteria for a Rigid Sediment Grain on a Rigid Surface}

\renewcommand{\thefootnote}{\fnsymbol{footnote}}

\author{S. J. Childs \footnotemark[1] \\ \\ {\small\em Department of Mechanical Engineering, Peninsula Technikon, Belville,} \\ {\small\em 7535, South Africa}}

\date{}       

\begin{document}
\maketitle
\footnotetext[1]{{\em schilds@iafrica.com}}
\renewcommand{\thefootnote}{\arabic{footnote}}

\begin{abstract}
\noindent {\em Criteria for the incipient motion of a rigid body initially
resting on a rigid surface are formulated from first principles in this
work. A modified Coulomb friction model and an associated distribution of
reaction forces are proposed. There exists a surprisingly large category
of general motions, however, which subscribe to a more conventional
analysis; an analysis made possible by identifying so--called
``significant reaction surfaces''. In this way a model which caters for
the majority of combined translations and rotations is devised. Some
introductry results demonstrate the accuracy with which fluids can be
numerically modelled for the purposes of entrainment. This work is an
extension of previous work by the same author.}
\end{abstract}


\section{Introduction}

There is nothing new about incipient motion analyses applied in a
sedimentary context (see Buffington and Montgomery (1997)). Although the
topic of sedimentation may not lend itself ideally to the application of
an incipient motion analysis, it is an ideal context in which to formulate
criteria for incipient motion, it being sufficiently general. 

A modified Coulomb friction model and an associated distribution of
reaction forces are proposed in this work. There exists a surprisingly
large category of general motions, however, which subscribe to a more
conventional analysis; an analysis which depends on identifying so--called
``significant reaction surfaces''. In this way a model which caters for
the majority of combined translations and rotations is devised.

The work presented in the subsequent sections is inextricably linked to the
theory and model developed by the author in {\sc Childs} (2000), {\sc
Childs} (1999) and {\sc Childs} and {\sc Reddy} (1999). This paper is a
supplement and concludes part of that work. The numerical approach
overcomes at least two of the limitations faced by experimentalists.
Localised intergranular flow is amenable to calculation (but not
measurement). What was previously an insurmountable problem in the
laboratory, reduces here to an exercise in mesh generation. Shape can be
more directly taken into account using numerical simulations and shape is
important. 

Incipient motion assumes selective erosion is as, or more, important than
selective deposition in an attempt to elucidate the problem of
sedimentation. It denies the possibility that some deposits may exist in a
state of dynamic equilibrium. 

\section{Using Numerical Methods to Model Entrainment Forces} \label{10}

\begin{figure}
\begin{center} \leavevmode
\mbox{\epsfbox{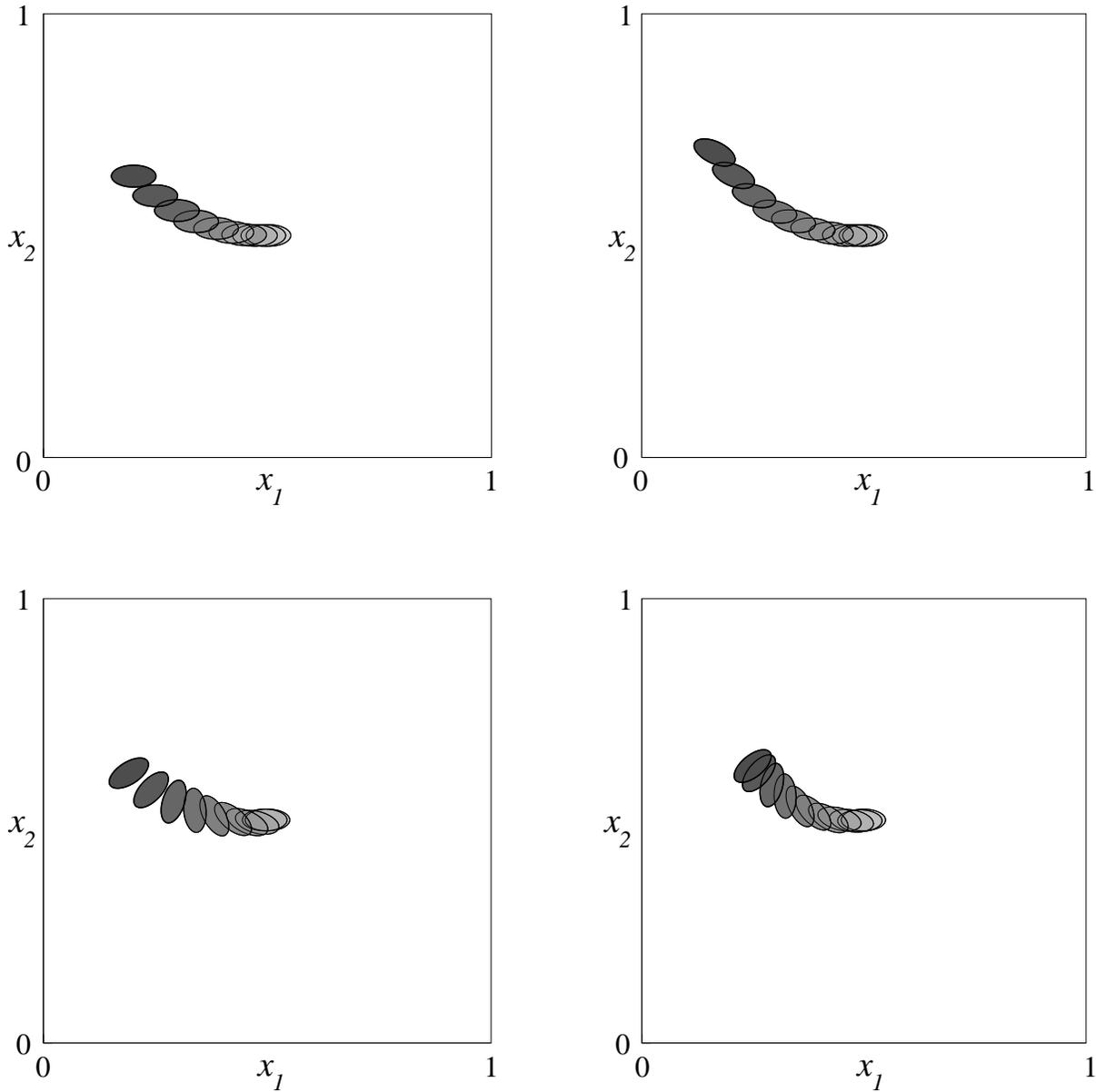}}
\end{center}
\caption{The trajectories of various included rigid bodies released
from rest at the centre of the driven cavity flow. {\sc Top
Left:} $Re = 0.025$, $m = 251.3$, $J_{33} = 314.2$ and
$t = 3.6$ secs. {\sc Top Right:} $Re = 0.025$, $m = 251.3$,
$J_{33} = 1.0$ and $t = 4.0$ secs. {\sc Bottom Left:} $Re =
0.025$, $m = 251.3$, $J_{33} = 0.1$ and $t = 3.6$ secs.
{\sc Bottom Right:} $Re = 1$, $m = 1$, moment of inertia
(scaled) $= 0.1 $ and $t = 2.0$ secs.} \label{108}
\end{figure}

\begin{figure}
\begin{center} \leavevmode
\mbox{\epsfbox{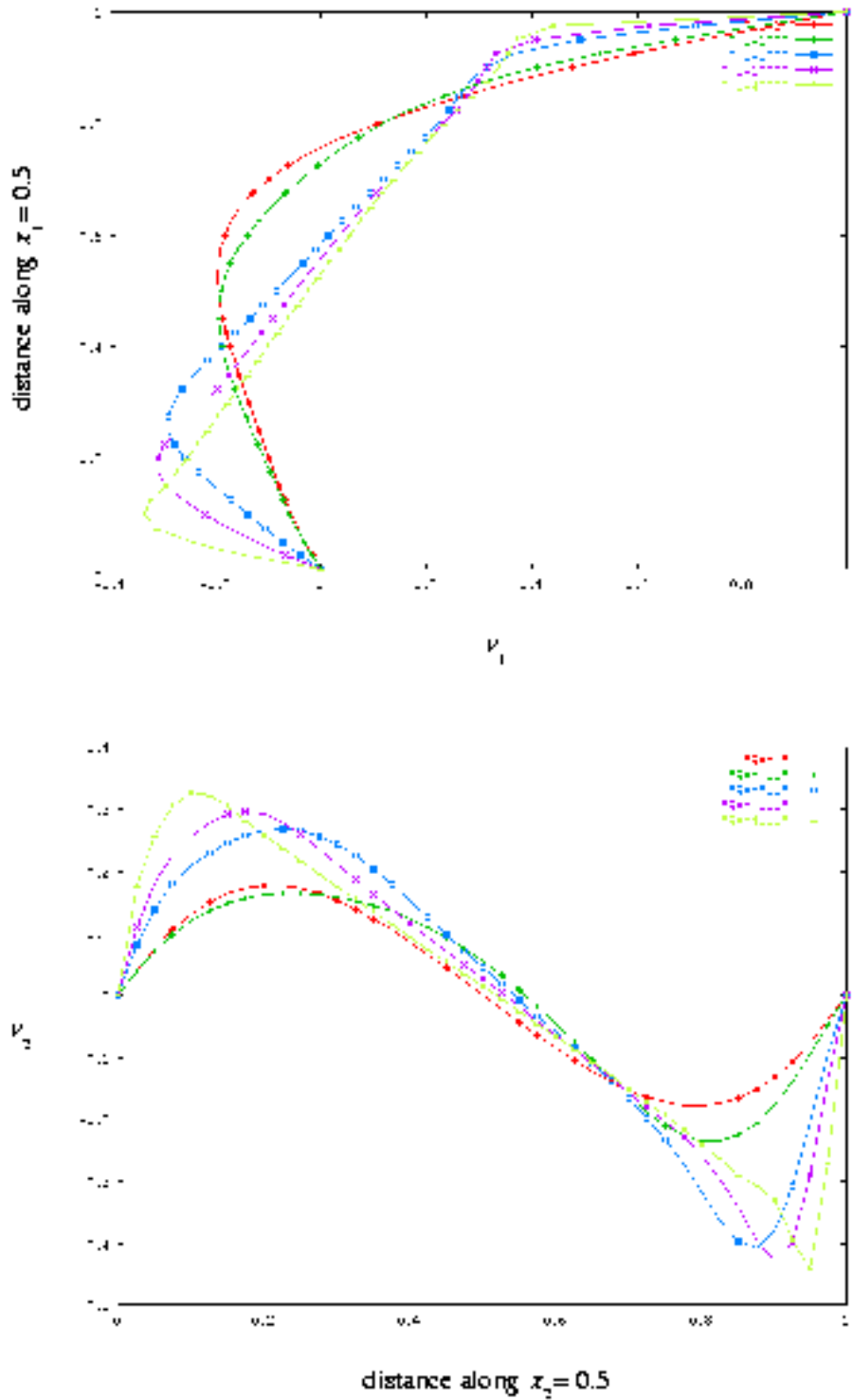}}
\end{center}
\caption{The respective $v_1$ and $v_2$ profiles computed along two cuts through the centre of a driven cavity flow (see {\sc Childs} (2000) and {\sc Childs} and {\sc Reddy} (1999)) for a range of Reynolds numbers up to 4000.} \label{882}
\end{figure}

The results for the ``pebble in a pothole'' (Fig. \ref{108}) are intended
to demonstrate something of the power with which state--of--the--art
numerical methods are able to model fluid flow and the forces exerted by
it. The ``pebble'' (a die bead of neutral
bouyancy\footnotemark[1]\footnotetext[1]{Although a neutral bouyancy is
not immediately reminiscent of any real life sediment problem, it was used
to give maximum meaning to the problem as a test.}) was released from rest
at the centre of the standard driven cavity flow problem (see Fig.
\ref{882}) and its motion was accurately determined. 

The combined, dimensionless, free surface--fluid--rigid body problem can
be stated as follows: Find ${\bx}_{rb}$, ${\btheta}$, ${\bv}_{rb}$,
${\bomega}$, $h$ and ${\bv}$ (the dimensionless, respective position,
orientation, velocity and angular velocity of the rigid body, the
elevation of the free surface and the velocity field of the fluid), which
satisfy  
\renewcommand{\thefootnote}{\fnsymbol{footnote}}
\footnotetext[2]{${\bF}={\bI}$, $J=1$ for correct implementations eg.
backward difference for time integration -- see {\sc Childs}
(2000)}
\begin{eqnarray}
J_{11} \displaystyle \frac{d \omega_{1}}{dt} \ +
\ (J_{33} \ - \ J_{22}) \omega_2 \omega_3 &=& \displaystyle \frac{\rho_{\scriptsize f}}{\rho_{\scriptsize s}} \left[ {\bH} \displaystyle \int_{\Gamma_{rb}} ({\bx}-{\bc} ) \wedge \left\{-{p}{\bI}
+ \frac{2}{Re} {{\bD}} \right\} {\bn} \ d{\Gamma}_{rb}
\right] \cdot {\be}_1 \nonumber \\
&& \label{331} \\ 
J_{22} \displaystyle \frac{d \omega_{2}}{dt} \ +
\ (J_{11} \ - \ J_{33}) \omega_3 \omega_1 &=& \displaystyle \frac{\rho_{\scriptsize
f}}{\rho_{\scriptsize s}} \left[ {\bH} \displaystyle \int_{\Gamma_{rb}} ({\bx}-{\bc}) \wedge \left\{-{p}{\bI}
+ \frac{2}{Re} {{\bD}} \right\} {\bn} \ d{\Gamma}_{rb}
\right] \cdot {\be}_2 \nonumber \\
&& \label{332} \\
J_{33} \displaystyle \frac{d \omega_{3}}{dt} \ +
\ (J_{22} \ - \ J_{11}) \omega_1 \omega_2 &=& \displaystyle \frac{\rho_{\scriptsize
f}}{\rho_{\scriptsize s}} \left[ {\bH} \displaystyle \int_{\Gamma_{rb}} ({\bx}-{\bc}) \wedge \left\{-{p}{\bI}
+ \frac{2}{Re} {{\bD}} \right\} {\bn} \ d{\Gamma}_{rb}
\right] \cdot {\be}_3 \nonumber \\
&& \label{333} \\ 
\displaystyle \frac{d{\bv}_{rb}}{dt} &=& \displaystyle \frac{\rho_{\scriptsize f}}{m
\rho_{\scriptsize s}} \int_{\Gamma_{rb}} \left\{ - {p}{\bI} +
\frac{2}{Re} {{\bD}} \right\} {\bn} \ d{\Gamma}_{rb} +
\frac{X}{V^2}{\bar {\bb}} \label{334} \\
&& \nonumber \\
\displaystyle \frac{d {{\btheta}}}{dt} &=&
{{\bomega}} \label{335} \\
&& \nonumber \\
\displaystyle \frac{d {\bx}_{rb}}{dt} &=&
{\bv}_{rb} \label{336} \\
&& \nonumber \\
\displaystyle \frac{\partial h}{\partial t} +
{\nabla h} \cdot [v_1, v_2] &=& v_3 \label{337} \\
&& \nonumber \\
\left[ \displaystyle \frac{\partial {\bv}}{\partial t} + {{\nabla {\bv} } {\bF}^{-1}
} ({\bv} - {\bv}^{mesh}) \right] J &=& \displaystyle
\frac{X}{V^2}{\bar {\bb}} J + \mathop{\rm div}{\bP} \hspace{5mm} \footnotemark[2] \label{338} \\
&& \nonumber \\
{ {\nabla {\bv} } : {\bF}^{-t} } &=& 0 \label{339}
\end{eqnarray}
\renewcommand{\thefootnote}{\arabic{footnote}}
where $\hspace{5mm} {\bP}  \ = \ {\bsigma} {\bF}^{-t}  J$ \\ \\
subject to the ``no slip'' requirements 
\begin{eqnarray*}
{\bv} \mid_{{\Gamma}_{rb}} &=& {\bv}_{rb} + {\bomega} \wedge ({\bx} - {\bc})
\end{eqnarray*}
at fluid--rigid body interfaces and
\begin{eqnarray*}
{\bv} \mid_{\Gamma} &=& {\bf 0} 
\end{eqnarray*}
at fixed, solid impermeable boundaries. Here $J_{ii \mbox{\scriptsize (no
sum)}}$ denotes the $i$th, dimensionless principal moment of inertia of
the rigid body, ${\bH}$ is the transition matrix for a transition to a
reference whose axes coincide with these principal moments of inertia,
${\bc}$ is the centre of mass of the rigid body, ${\Gamma}_{rb}$ is the
dimensionless surface of the rigid body, $t$ is a dimensionless time,
$\rho_f$ and $\rho_s$ are the fluid density and units of solid density
respectively, $m$ is the dimensionless mass of the rigid body,
$\frac{X}{V^2}{\bar {\bb}}$ is a dimensionless body force per unit mass,
${\bF}$ is the deformation gradient, $J$ its determinant,
${\bv}^{\scriptsize \mbox{\it mesh}}$ is the dimensionless velocity of a
reference which is otherwise allowed to deform freely and $Re$ is the
Reynolds number. The Cauchy stress, which appears in the formula for
${\bP}$ (the Piola--Kirchoff stress tensor of the first kind) is that
given in terms of the constitutive relation for a Newtonian fluid,
$\bsigma = - p {\bI} + 2 \mu {\bD}$. A fully viscous boundary layer is
therefore an intrinsic property of this model. All quantities otherwise
undefined have their usual meanings.

The first four equations (Eqns. (\ref{331}) -- (\ref{334})) govern the
motion of a rigid body in a fluid. Eqns. (\ref{331}), (\ref{332}) and
(\ref{333}) are a dimensionless form of Euler's equations written in terms
of a flow tractional force. They embody conservation of angular momentum.
Eqn. (\ref{334}) embodies conservation of linear momentum. Eqns.
(\ref{335}) and (\ref{336}) facilitate recovery of the solution ultimately
sought, the orientation and position (as opposed to angular and linear
velocity). Eqn. (\ref{337}) is the free surface model, one in which
surface tension is considered negligeable. Eqns. (\ref{338}) and
(\ref{339}) are a completely general reference description of a viscous,
incompressible fluid written in dimensionless form. It should, however, be
noted that $\bv$ is merely a function of a distorting reference and the
description therefore remains inertial in the same way that Lagrangian
descriptions do. Both fluid and free surface sub--problems are best solved
using the finite element method. A fourth order Runge--Kutta--Fehlberg
method was used to solve the entire, coupled rigid body sub--problem.
Rigid body, free surface and fluid sub--problems were solved iteratively
thereby enforcing the entire set of governing equations. 

A comprehensive derivation of all equations can be found in {\sc Childs}
(2000), {\sc Childs} (1999) and {\sc Childs} and {\sc Reddy} (1999). Most
of the numerical details, particularly those pertaining to the finite
element method, can be found in the latter. Of course turbulence is not
taken into account in this particular model. It is possible to incorporate
turbulent features by way of modifying the viscosity according to a well
established technique, the so--called $k$--$\epsilon$ model. The model
presented here may therefore be considered to have certain limitations.   

In the succesive Fig. \ref{108} trajectories the mass was concentrated
closer to the centre (a lower moment of inertia was used). The Reynolds
number used was based on grain size in all but the bottom right set of
results in Fig. \ref{108}. In this instance the so-called flow--field
Reynolds number was used. 

\section{Incipient Motion}

The possibility of mobilisation (or remobilisation) is, for sediments,
perhaps more relevant than the transport of the rigid body by the fluid.
Incipient motion can be used as a simplistic criterion on which to base
deposition, consequently the hydrodynamic characterisation of sediments
and their environments of deposition. 

\subsection{Notation}

\begin{figure}
\begin{center} \leavevmode
\mbox{\epsfbox{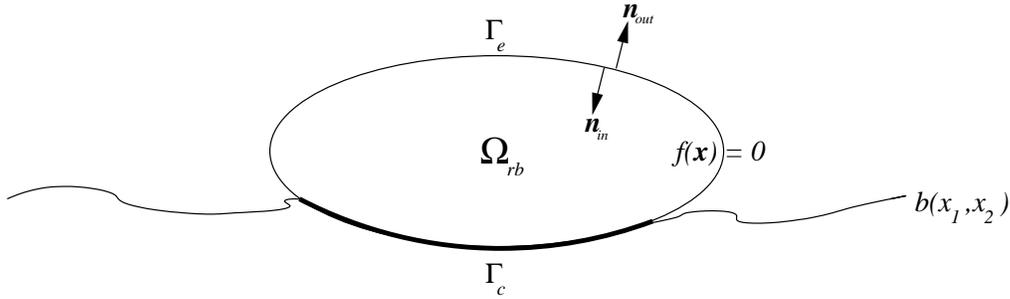}}
\end{center}
\caption{${\Gamma}_e$ and ${\Gamma}_c$ are the respective exposed and
contact surfaces of a rigid body of shape $f({\bx}) = 0$, resting on a
rigid bottom of shape $b(x_1, x_2)$.} \label{188}
\end{figure}

Suppose that the shape of the rigid body can be adequately described by
some equation of the form
\[
f({\bx}) = 0
\]
and that the rigid surface on which this rigid body rests can be
adequately described in terms of some function $b(x_1, x_2)$ which specifies the vertical position of the surface in terms of the horizontal coordinates $x_1$ and $x_2$. That is
\[
x_3 = b(x_1, x_2)
\]
(see Fig. \ref{188} for a schematic representation of the problem of
interest).
A contact surface and an exposed surface, $\Gamma_c$ and $\Gamma_e$ respectively, may be formally defined as
\[
\Gamma_c = \left\{ {\bx} : f({\bx}) = 0, x_3 = b(x_1, x_2); {\bx} \in
{\cal R} \right\}
\]
and
\[
\Gamma_e = \left\{ {\bx} : f({\bx}) = 0, x_3 \not= b(x_1, x_2); {\bx}
\in {\cal R} \right\}
\]
in terms of this notation. The surface of the rigid body has outward and inward normals, $\bn_{\mbox{\scriptsize \it out}}$ and $\bn_{\mbox{\scriptsize \it in}}$ respectively, 
\[
\bn_{\mbox{\scriptsize \it out}} = - \bn_{\mbox{\scriptsize \it in}} = \frac{\nabla f}{\mid\mid \nabla f \mid\mid}.
\]
One might expect $f({\bx})$ and $b(x_1, x_2)$ to be complicated in real--life examples. 

\subsection{Modelling Incipient Motion}

A more direct application of Newton's first law one possibly couldn't find
than incipient motion. The force acting on a rigid sediment grain is best thought of as
\begin{eqnarray*}
{\bF} = {\bF}_{\mbox{\scriptsize \it flow\&${\bg}$}} + {\bF}_c
\end{eqnarray*}
where ${\bF}_{\mbox{\scriptsize \it flow\&${\bg}$}}$ is the combined
gravitational and flow--tractional force acting on the rigid body and
${\bF}_c$ is the sum of forces (reaction and frictional) acting at the
bed--rigid body contact, $\Gamma_c$. Any incipient translation will be
initiated by a combination of gravitational and flow--tractional forces,
\begin{eqnarray} \label{9}
{\bF}_{\mbox{\scriptsize \it flow\&${\bg}$}} = \int_{\Gamma_e} {\bsigma} \mbox{$\bn_{\mbox{\scriptsize \it in}}$} d\Gamma_e + m{\bg},
\end{eqnarray}
modified by any induced forces of reaction and friction ($\bsigma$ is the
Cauchy stress at $\Gamma_e$, the rigid body--fluid interface, and m{\bg}
is the weight of the particle). The combined gravitational and flow--tractional couple is
\begin{eqnarray} \label{21}
{\btau}_{\mbox{\scriptsize \it flow\&${\bg}$}} &=& \int_{\Gamma_e} ({\bx} - {\bc}) \wedge {\bsigma} \mbox{$\bn_{\mbox{\scriptsize \it in}}$} \ d\Gamma_e + \int_{\Omega_{rb}} \rho ({\bx} - {\bc}) \wedge {\bg} \ d\Omega
\end{eqnarray}
where ${\bc}$ is some convenient point\footnotemark[2] and $\Omega_{rb}$ is the rigid body domain.\footnotetext[2]{The torque is independent of the point about which it was calculated}

The important thing to recognise here is that any incipient translation
will nonetheless be a component of the force ${\bF}_{\mbox{\scriptsize \it
flow\&${\bg}$}}$ (other components contributing to couples or being
cancelled outright) and any incipient rotation will likewise be a
component of ${\btau}_{\mbox{\scriptsize \it flow\&${\bg}$}}$.

A modified or extended Coulomb friction model can be arrived at by
extending the model for rigid, planar contact surfaces to more general surfaces i.e. the distribution of the reaction forces along $\Gamma_c$ is obtained by similar use of Newton's third law. The reaction force for a plane contact surface is 
\begin{eqnarray} \label{31}
{\bF}_{\mbox{\scriptsize \it reaction}} &=& \left( \bn_{\mbox{\scriptsize
\it out}} \cdot {\bF}_{\mbox{\scriptsize \it flow\&${\bg}$}} \right)
\bn_{\mbox{\scriptsize \it in}}
\end{eqnarray}
and the associated frictional force is
\begin{eqnarray} \label{33}
{\bF}_{\mbox{\scriptsize \it friction}} &=& - C_{\mbox{\scriptsize \it friction}} \left[{\bF}_{\mbox{\scriptsize \it flow\&${\bg}$}} - \left( \bn_{\mbox{\scriptsize \it out}} \cdot {\bF}_{\mbox{\scriptsize \it flow\&${\bg}$}} \right) \bn_{\mbox{\scriptsize \it out}}\right]
\end{eqnarray}
where $C_{\mbox{\scriptsize \it friction}} = \min \left\{ 1, \frac{\zeta 
\left( \bn_{\mbox{\scriptsize \it out}} \cdot {\bF}_{\mbox{\scriptsize \it
flow\&${\bg}$}} \right)}{\left|\left| {\bF}_{\mbox{\scriptsize \it
flow\&${\bg}$}} - \left( \bn_{\mbox{\scriptsize \it out}} \cdot
{\bF}_{\mbox{\scriptsize \it flow\&${\bg}$}} \right)
\bn_{\mbox{\scriptsize \it out}} \right|\right|} \right\}$ and $\zeta$ is
the coefficient of static friction. Extending the above models to more
general surfaces
\begin{eqnarray*}
{\bF}_c &=& c_1 \displaystyle \int_{\Gamma_c}
\left({\bF}_{\mbox{\scriptsize \it reaction}} + {\bF}_{\mbox{\scriptsize
\it friction}}\right) H\left(\bn_{\mbox{\scriptsize \it out}} \cdot
{\bF}_{\mbox{\scriptsize \it flow\&${\bg}$}}\right) d{\Gamma_c} \nonumber 
\end{eqnarray*}
is obtained where $H$ is the Heaviside step function and $c_1$ is some scaling factor
\begin{eqnarray*}
\begin{array}{c} c_1 = \\ \\ \end{array} \underbrace{\begin{array}{c}
\displaystyle \frac{c_2}{\int_{\Gamma_c} H\left(\bn_{\mbox{\scriptsize \it out}} \cdot {\bF}_{\mbox{\scriptsize \it
flow\&${\bg}$}}\right) d{\Gamma_c}} \\ \\ \end{array}}_{
\mbox{\scriptsize $c_2 \ /$ the pertinent area}} 
\end{eqnarray*}
in which $c_2$ is determined by requiring that a component of
${\bF}_{\mbox{\scriptsize \it flow\&${\bg}$}}$, normal to some part of the
contact, be completely balanced by the same component of reaction force.
The disturbing question of whether another such component will be likewise
balanced, when using the same $c_2$ scale, then arises. This dilemma can
be avoided for all but one category of incipient motion in the forthcoming
analysis. 

For couples,
\begin{eqnarray*}
{\btau}_{c} &=& c_1 \displaystyle \int_{\Gamma_c} \left[({\bx} -
{\bc}) \wedge {\bF}_{\mbox{\scriptsize \it reaction}} + ({\bx} -
{\bc}) \wedge {\bF}_{\mbox{\scriptsize \it friction}}\right]
H\left(\mbox{$\bn_{\mbox{\scriptsize \it out}}$} \cdot
{\bF}_{\mbox{\scriptsize \it flow\&${\bg}$}}\right) \ d{\Gamma_c}.
\end{eqnarray*}
There exists a surprisingly large category of general motions which
subscribe to a more conventional analysis than the one just proposed. 

Possible modes of incipient motion are categorised as follows for the purposes of this work:
\begin{enumerate}
\item \label{11} Translation only (lifting or sliding).
\item \label{12} Sliding combined with an ``away from the surface'' rotation. 
\item \label{13} Lifting combined with an ``away from the surface'' rotation.
\item \label{14} Pivotting combined with lifting.
\item \label{15} Rotation only (pivotting or an ``away from the surface'' rotation).
\item \label{16} Pivotting combined with sliding: \begin{enumerate}
\item \label{17} About the same contact.
\item \label{18} About different contacts (less likely).
\end{enumerate}
\end{enumerate}
By ``away from the surface'' rotations is meant that the sediment partcle
is rotated off the contact surface in such a way that no frictional or
reaction forces are incurred i.e. there is no counter torque. This condition amounts (under all but the most exceptional circumstances)
to immediately excluding axes of rotation surrounded and intersected by
normals from the contact surface. 
\begin{figure}
\begin{center} \leavevmode
\mbox{\epsfbox{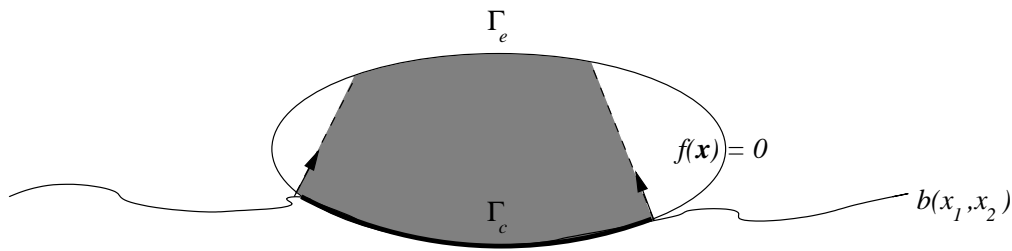}}
\end{center}
\caption{The condition which amounts to immediately excluding axes of
rotation around which normals to the contact surface can be drawn.}
\label{191}
\end{figure}
In Fig. \ref{191} the axis about which an ``away from the surface''
rotation occurs lies in one of the unshaded regions -- which one depends
on the sign of the rotation. For flow--induced axes of rotation which lie
in a region such as the shaded one in Fig. \ref{191}, one might presume
the point about which the rigid body will rotate, is the nearest point to
the centre of mass which does not lie on a normal to the contact surface
(based on a principle of least action).

\subsection{Incipient Translation Only (${\btau}_{\mbox{\scriptsize \it flow\&${\bg}$}} = \mbox{\bf0}$)}

\begin{figure}[H]
\begin{center} \leavevmode
\mbox{\epsfbox{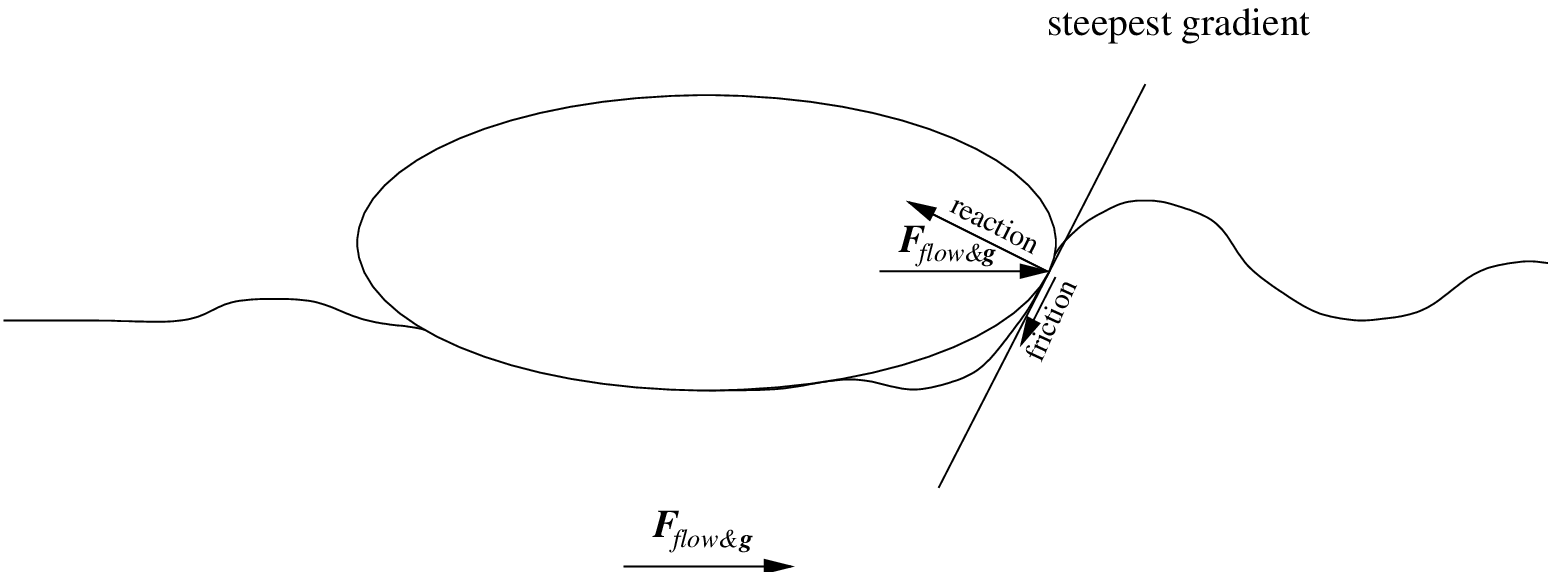}}
\end{center}
\caption{The significant reaction surface to ${\bF}_{\mbox{\scriptsize \it flow\&${\bg}$}}$.} \label{189}
\end{figure}
The significant reaction surface is that part of the contact surface which determines the direction of the incipient translation, here 
\begin{eqnarray} \label{2}
\min_{{\bx} \in \Gamma_c} \mbox{$\bn_{\mbox{\scriptsize \it in}}$} \cdot {\bF}_{\mbox{\scriptsize \it flow\&${\bg}$}}. 
\end{eqnarray}
To work out the nett or resultant force compute ${\bF}_{\mbox{\scriptsize
\it flow\&${\bg}$}}$ (Eqn. (\ref{9})), then locate the significant reaction
surface\footnotemark[3]\footnotetext[3]{assumed unique for simplicity,
otherwise the forces will be equally shared} in $\Gamma_c$ (Eqn.
(\ref{2})). The resultant force which arises is
\begin{eqnarray} \label{3}
{\bF} &=& {\bF}_{\mbox{\scriptsize \it flow\&${\bg}$}} + \left({\bF}_{\mbox{\scriptsize \it reaction}} + {\bF}_{\mbox{\scriptsize \it friction}}\right) H\left(\mbox{$\bn_{\mbox{\scriptsize \it out}}$} \cdot {\bF}_{\mbox{\scriptsize \it flow\&${\bg}$}}\right)
\end{eqnarray}
where ${\bF}_{\mbox{\scriptsize \it reaction}}$ and ${\bF}_{\mbox{\scriptsize \it friction}}$ are given by equations (\ref{31}) and (\ref{33}) respectively. The resulting torque is
\begin{eqnarray*}
{\btau} &=& \left[({\bx}^* - {\bc}) \wedge {\bF}_{\mbox{\scriptsize
\it reaction}} + ({\bx}^* - {\bc}) \wedge {\bF}_{\mbox{\scriptsize
\it friction}}\right] H\left(\bn_{\mbox{\scriptsize \it out}} \cdot {\bF}_{\mbox{\scriptsize \it flow\&${\bg}$}}\right)
\end{eqnarray*}
where ${\bx}^*$ is the minimum prescribed by Eqn. (\ref{2}). 

\subsection{Incipient Rotation Only (${\bF}_{\mbox{\scriptsize \it flow\&${\bg}$}} = \mbox{\bf0}$)}

One can surmise that the torque ${\btau}_{\mbox{\scriptsize \it flow\&${\bg}$}}$ is equivalent to a tangential force
\begin{eqnarray} \label{6}
{\bF}_{\mbox{\scriptsize \it ${\btau}$}} = \frac{{\btau}_{\mbox{\scriptsize \it flow\&${\bg}$}} \wedge ({\bx} - {\bc})}{\mid\mid {\bx} - {\bc} \mid\mid^2},
\end{eqnarray}
acting at ${\bx}$, since ${\btau}_{\mbox{\scriptsize \it
flow\&${\bg}$}} = ({\bx} - {\bc}) \wedge {\bF}_{\mbox{\scriptsize \it
${\btau}$}}$ by definition. The counter--couple arising due to a
reaction at the contact surface has a negative projection on the
combined gravitational and flow--tractional torque at the desired
location, furthermore, it is identified by the value of this
projection being a minimum. That is,
\begin{eqnarray} \label{7}
\min_{{\bx} \in \Gamma_c} \left\{ \left[ \left({\bx} -
{\bc}\right) \wedge \left( \frac{{\btau}_{\mbox{\scriptsize \it flow\&${\bg}$}} \wedge ({\bx} - {\bc})}{\mid\mid {\bx} - {\bc} \mid\mid^2} \cdot \bn_{\mbox{\scriptsize \it in}} \right) \right] \cdot {\btau}_{\mbox{\scriptsize \it flow\&${\bg}$}} \right\}.
\end{eqnarray}
\begin{figure}
\begin{center} \leavevmode
\mbox{\epsfbox{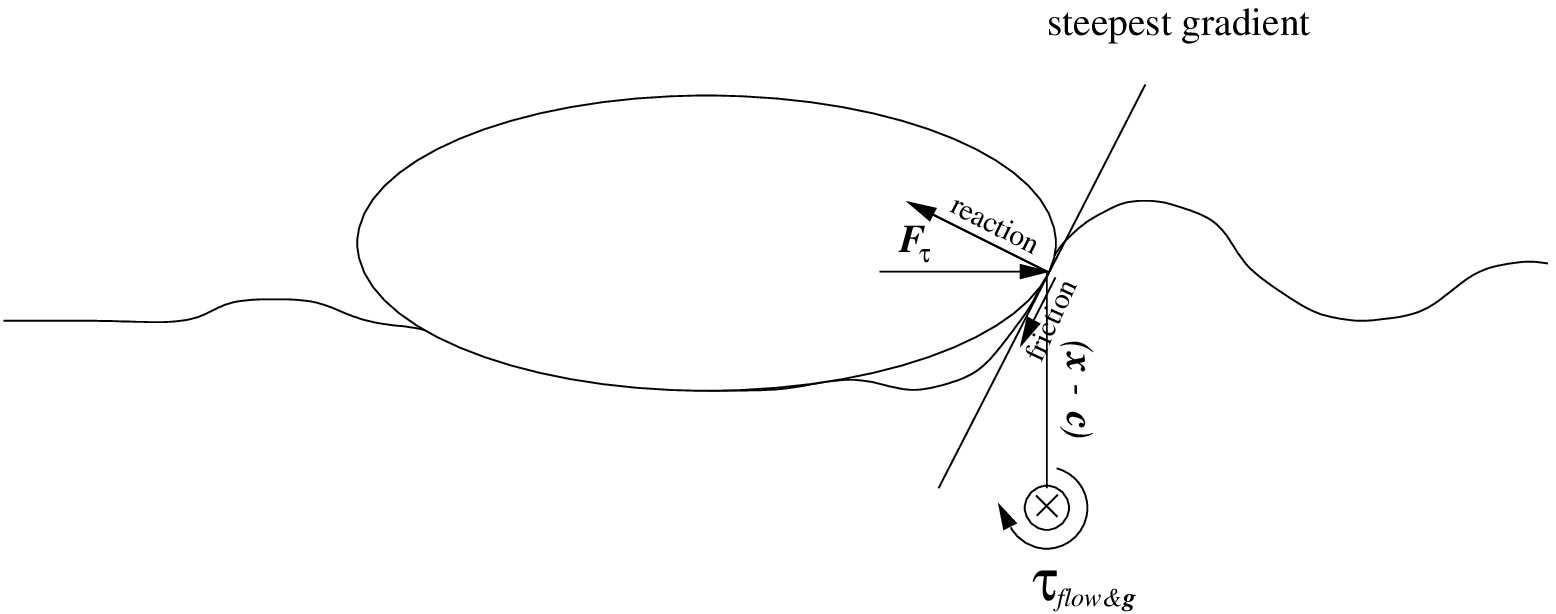}}
\end{center}
\caption{The significant reaction surface to ${\btau}_{\mbox{\scriptsize \it flow\&${\bg}$}}$.} \label{190}
\end{figure}
A strategy for computing the torque exerted on the rigid body may
therefore be summarised as follows. Compute ${\btau}_{\mbox{\scriptsize
\it flow\&${\bg}$}}$ (using Eqn. (\ref{21})) then locate the position of the significant reaction surface in $\Gamma_c$ (using Eqn. (\ref{7})). The resultant torque which arises (assuming the minimum is unique for simplicity) can then be calculated using the formula
\begin{eqnarray} \label{8}
{\btau} &=& {\btau}_{\mbox{\scriptsize \it flow\&${\bg}$}} + \left\{
\mbox{$\left( \bn_{\mbox{\scriptsize \it out}} \cdot
{\bF}_{\mbox{\scriptsize \it ${\btau}$}} \right) \left({\bx} -
{\bc}\right) \wedge \bn_{\mbox{\scriptsize \it in}}$} \right. \nonumber
\\  && \left. - C_{\mbox{\scriptsize \it friction}}  \left({\bx} -
{\bc}\right) \wedge \left[{\bF}_{\mbox{\scriptsize \it ${\btau}$}} -
\left( \bn_{\mbox{\scriptsize \it out}} \cdot {\bF}_{\mbox{\scriptsize \it
${\btau}$}} \right) \bn_{\mbox{\scriptsize \it out}}\right] \right\}
H\left(\mbox{$\bn_{\mbox{\scriptsize \it out}}$} \cdot
{\bF}_{\mbox{\scriptsize \it ${\btau}$}}\right)
\end{eqnarray}
where $C_{\mbox{\scriptsize \it friction}} = \min \left\{ 1, \frac{\zeta \left( \bn_{\mbox{\scriptsize \it out}} \cdot {\bF}_{\mbox{\scriptsize \it ${\btau}$}} \right)}{\left|\left| {\bF}_{\mbox{\scriptsize \it ${\btau}$}} - \left( \bn_{\mbox{\scriptsize \it out}} \cdot {\bF}_{\mbox{\scriptsize \it ${\btau}$}} \right) \bn_{\mbox{\scriptsize \it out}} \right|\right|} \right\}.$ Both frictional and reactional torques are drawn up along similar lines to ${\bF}_{\mbox{\scriptsize \it friction}}$ and ${\bF}_{\mbox{\scriptsize \it reaction}}$ were. A non--zero value indicates the frictional force is insufficient to impede rotation. The translational force (a result of any uncoupling) is
\begin{eqnarray*}
{\bF} = \left\{ \left( \bn_{\mbox{\scriptsize \it out}} \cdot {\bF}_{\mbox{\scriptsize \it ${\btau}$}} \right) \bn_{\mbox{\scriptsize \it in}} - C_{\mbox{\scriptsize \it friction}} \ \left[{\bF}_{\mbox{\scriptsize \it ${\btau}$}} - \left( \bn_{\mbox{\scriptsize \it out}} \cdot {\bF}_{\mbox{\scriptsize \it ${\btau}$}} \right) \bn_{\mbox{\scriptsize \it out}}\right] \right\} H\left(\bn_{\mbox{\scriptsize \it out}} \cdot {\bF}_{\mbox{\scriptsize \it ${\btau}$}}\right).
\end{eqnarray*}

\subsection{Simultaneous Incipient Rotation and Translation (${\bF}_{\mbox{\scriptsize \it flow\&${\bg}$}} \neq \mbox{\bf0}$, ${\btau}_{\mbox{\scriptsize \it flow\&${\bg}$}} \neq \mbox{\bf0}$)}

Assuming a common position for both the pivot and the initial sliding
surface, it can be located using either the (\ref{2}) or (\ref{7}) minima.
To determine the nett or resultant force and torque, compute
${\bF}_{\mbox{\scriptsize \it flow\&${\bg}$}}$ and
${\btau}_{\mbox{\scriptsize \it flow\&${\bg}$}}$ (using equations
(\ref{9}) and (\ref{21})). Locate the position on $\Gamma_c$ at which
either of the minima, (\ref{2}) or (\ref{7}), occur. At this point
evaluate the quantities ${\bF}_{\mbox{\scriptsize \it ${\btau}$}}$ (using
Eqn. (\ref{6})),
\begin{eqnarray*}
{\bF}_{\mbox{\scriptsize \it flow${\bg}$\&${\btau}$}} &=&
{\bF}_{\mbox{\scriptsize \it flow\&${\bg}$}} + {\bF}_{\mbox{\scriptsize \it ${\btau}$}}\mid_{{\bx}^*},
\end{eqnarray*}
\begin{eqnarray} \label{22}
\mbox{$\bn_{\mbox{\scriptsize \it in}}$} \cdot {\bF}_{\mbox{\scriptsize \it flow${\bg}$\&${\btau}$}} \mid_{{\bx}^*} 
\end{eqnarray}
and
\begin{eqnarray} \label{23}
\left[ \left({\bx} - {\bc}\right)
\wedge \left( {\bF}_{\mbox{\scriptsize \it flow${\bg}$\&${\btau}$}} \cdot \bn_{\mbox{\scriptsize \it in}} \right) \right] \cdot
{\btau}_{\mbox{\scriptsize \it flow\&${\bg}$}} \mid_{{\bx}^*} 
\end{eqnarray}
where ${\bx}^*$ is the prescribed minimum. The resultant force which arises in the event of Eqn. (\ref{22}) being a negative quantity can be calculated according to
\begin{eqnarray} \label{24}
{\bF} &=& {\bF}_{\mbox{\scriptsize \it flow\&${\bg}$}} + \left\{ \mbox{$\left( \bn_{\mbox{\scriptsize \it out}} \cdot {\bF}_{\mbox{\scriptsize \it flow${\bg}$\&${\btau}$}} \right) \ \bn_{\mbox{\scriptsize \it in}}$}  \right. \nonumber \\ 
&& \left. - C_{\mbox{\scriptsize \it friction}} \ \left[{\bF}_{\mbox{\scriptsize \it flow${\bg}$\&${\btau}$}} - \left( \bn_{\mbox{\scriptsize \it out}} \cdot {\bF}_{\mbox{\scriptsize \it flow${\bg}$\&${\btau}$}} \right) \bn_{\mbox{\scriptsize \it out}}\right] \right\} H\left(\mbox{$\bn_{\mbox{\scriptsize \it out}}$} \cdot {\bF}_{\mbox{\scriptsize \it flow${\bg}$\&${\btau}$}}\right) 
\end{eqnarray}
where $C_{\mbox{\scriptsize \it friction}} = \min \left\{ 1, \frac{\zeta \left( \bn_{\mbox{\scriptsize \it out}} \cdot {\bF}_{\mbox{\scriptsize \it flow${\bg}$\&${\btau}$}} \right)}{\left|\left| {\bF}_{\mbox{\scriptsize \it flow${\bg}$\&${\btau}$}} - \left( \bn_{\mbox{\scriptsize \it out}} \cdot {\bF}_{\mbox{\scriptsize \it flow${\bg}$\&${\btau}$}} \right) \bn_{\mbox{\scriptsize \it out}} \right|\right|} \right\}.$ A zero value would correspond to no incipient motion. A non--zero value indicates the frictional force is insufficient to impede translation. The resulting torque is
\begin{eqnarray} \label{25}
{\btau} &=& {\btau}_{\mbox{\scriptsize \it flow\&${\bg}$}} + \left\{ \mbox{$\left( \bn_{\mbox{\scriptsize \it out}} \cdot {\bF}_{\mbox{\scriptsize \it flow${\bg}$\&${\btau}$}} \right) ({\bx}^* - {\bc}) \wedge \bn_{\mbox{\scriptsize \it in}}$} \right. \nonumber \\  
&& \left. - C_{\mbox{\scriptsize \it friction}} \ \left[{\bF}_{\mbox{\scriptsize \it flow${\bg}$\&${\btau}$}} - \left( \bn_{\mbox{\scriptsize \it out}} \cdot {\bF}_{\mbox{\scriptsize \it flow${\bg}$\&${\btau}$}} \right) \bn_{\mbox{\scriptsize \it out}}\right]  \right\} H\left(\mbox{$\bn_{\mbox{\scriptsize \it out}}$} \cdot {\bF}_{\mbox{\scriptsize \it flow${\bg}$\&${\btau}$}}\right) 
\end{eqnarray}
where ${\bx}^*$ is the prescribed minimum. The outcome could be summarised as follows:
\begin{enumerate}
\item ${\bF} \neq \mbox{\bf0}$ by Eqn. (\ref{24}) and ${\btau} = \mbox{\bf0}$ by Eqn. (\ref{25}) $\Rightarrow$ translation.
\begin{enumerate}
\item minimum (\ref{2}) $< 0 \Rightarrow$ sliding
\item minimum (\ref{2}) $> 0 \Rightarrow$ lifting.
\end{enumerate}
\item Minimum (\ref{2}) $< 0$, minimum (\ref{7}) $> 0 \Rightarrow$ sliding with an ``away from the surface'' rotation.
\item Minimum (\ref{2}) $> 0$, minimum (\ref{7}) $> 0 \Rightarrow$ lifting with an ``away from the surface'' rotation.
\item Minimum (\ref{2}) $> 0$, minimum (\ref{7}) $< 0 \Rightarrow$ pivotting.
\item Minima (\ref{2}) and (\ref{7}) $< 0$, ${\bF} = \mbox{\bf0}$ by Eqn. (\ref{24}) and ${\btau} \neq \mbox{\bf0}$ by Eqn. (\ref{25}) $\Rightarrow$ rotation only.
\begin{enumerate}
\item minimum (\ref{7}) $< 0 \Rightarrow$ pivotting
\item minimum (\ref{7}) $> 0 \Rightarrow$ ``away from the surface'' rotation.
\end{enumerate}
\item Minima (\ref{2}) and (\ref{7})$< 0$, ${\bF} \neq \mbox{\bf0}$ by Eqn. (\ref{24}) and ${\btau} \neq \mbox{\bf0}$ by Eqn. (\ref{25}) $\Rightarrow$ sliding and pivotting.
\item ${\bF} = \mbox{\bf0}$ by Eqn. (\ref{24}) and ${\btau} = \mbox{\bf0}$ by Eqn. (\ref{25}) $\Rightarrow$ no incipient motion
\end{enumerate}
Note that ``pivotting and sliding about different (less likely) contacts''
is the only mode not comprehensively dealt with. 

\section{Conclusions}

Classes of permissible incipient rotations and translations can be readily
and systematically formulated for rigid bodies placed on rigid surfaces,
as can be a modified Coulomb friction model. These classes of permissible
incipient motion together with the friction models, are proposed as
criteria on which to base incipient translation and rotation.

The possibility of mobilisation (or remobilisation) is, for sediments,
perhaps more relevant than the transport of the rigid body by the fluid. Incipient motion can be used as a simplistic criterion on which to base
deposition, consequently the hydrodynamic characterisation of sediments
and their environments of deposition. The unknown final dynamic or static
state of the depositional equilibrium could, in some instances, have
serious ramifications for incipient motion based sedimentation models.

\section{Acknowledgements}
 
Frank Shillington is acknowledged for the considerable support and
motivation which he provided at the time when much of this work was
initially done. Graeme Oliver is thanked for generously funding the author
from his research grant. 

\nocite{me:3}
\nocite{me:1}
\nocite{me:2}
\nocite{oden:3}
\nocite{buffington:2}

\bibliography{incipient}

\end{document}